\documentstyle[12pt]{article}
\date{}
\author{Valerii Dryuma\thanks{Work supported in part by Grant RFFI, Russia-Moldova}\\[5mm]
{\it Institute of Mathematics and Informatics, AS RM,}\\[3mm] {\it
5 Academiei Street, 2028 Kishinev, Moldova},\\[3mm]{\it e-mail:
valery@dryuma.com; cainar@mail.md} }
\title{On geometry of gonometric family of cycles}

\newtheorem{pr}{Proposition}
\newtheorem{rem}{Remark}
\begin{document}
\maketitle
\date{}
\maketitle
\begin{abstract}
\ \ \ \ An examples solutions of the equation for curvature of
congruence of cycles  are considered. Their properties are
discussed.
\end{abstract}


\medskip
\section{Gonometric family of cycles}

      Two parametrical family of cycles on the plane is determined by the equation
      \[
      (\xi-x)^2+(\eta-y)^2-\phi(x,y)^2=0.
      \]

     The angle metric in a given family of cycles is defined by the
     expression \cite{sur}
     \begin{equation}\label{dr:eq1}
     {{\it ds}}^{2}={\frac {\left (1-\left ({\frac {\partial }{\partial x}}
\phi(x,y)\right )^{2}\right ){{\it dx}}^{2}-2\,{\frac {
\partial }{\partial x}}\phi(x,y){\frac {\partial }{
\partial y}}\phi(x,y){\it dx}\,{\it dy}+\left (1-\left ({
\frac {\partial }{\partial y}}\phi(x,y)\right )^{2}\right ){{\it
dy}}^ {2}}{\left (\phi(x,y)\right )^{2}}} .
 \end{equation}

     The expression for the curvature of the metric (\ref{dr:eq1}) has the form

\begin{equation}\label{dr:eq2}
K(x,y)={\frac {\phi\left (\left (1-\left ({\frac {\partial
}{\partial y} }\phi\right )^{2}\right ){\frac {\partial
^{2}}{\partial {x}^{2}} }\phi+2\,\left ({\frac {\partial
}{\partial y}}\phi\right ) \left ({\frac {\partial }{\partial
x}}\phi\right ){\frac {
\partial ^{2}}{\partial x\partial y}}\phi+\left (1-\left ({\frac
{\partial }{\partial x}}\phi\right )^{2}\right ){\frac {\partial
^{2}}{\partial {y}^{2}}}\phi\right )}{\left (1-\left ({\frac {
\partial }{\partial x}}\phi\right )^{2}-\left ({\frac {\partial }
{\partial y}}\phi\right )^{2}\right )^{2}}}-\]\[-{\frac
{\phi^{2}\left (\left ({\frac {\partial ^{2}}{\partial {x}^{2}}}
\phi\right ){\frac {\partial ^{2}}{\partial {y}^{2}}}\phi- \left
({\frac {\partial ^{2}}{\partial x\partial y}}\phi\right )^
{2}\right )}{\left (1-\left ({\frac {\partial }{\partial x}}\phi
\right )^{2}-\left ({\frac {\partial }{\partial y}}\phi\right )^{
2}\right )^{2}}}-\left (1-\left ({\frac {\partial }{\partial
x}}\phi\right )^{2}-\left ({\frac {\partial }{\partial
y}}\phi\right )^{2}\right )^{-1}+1.
\end{equation}

\medskip
\section{The congruence of cycles of constant curvature}

      The congruence of cycles of constant curvature $K(x,y)=K$
      are defined by the Monge-Ampere equation \cite{sur}
\begin{equation}\label{dr:eq3}
K={\frac {\phi\left (\left (1-\left ({\frac {\partial }{\partial
y} }\phi\right )^{2}\right ){\frac {\partial ^{2}}{\partial
{x}^{2}} }\phi+2\,\left ({\frac {\partial }{\partial y}}\phi\right
) \left ({\frac {\partial }{\partial x}}\phi\right ){\frac {
\partial ^{2}}{\partial x\partial y}}\phi+\left (1-\left ({\frac
{\partial }{\partial x}}\phi\right )^{2}\right ){\frac {\partial
^{2}}{\partial {y}^{2}}}\phi\right )}{\left (1-\left ({\frac {
\partial }{\partial x}}\phi\right )^{2}-\left ({\frac {\partial }
{\partial y}}\phi\right )^{2}\right )^{2}}}-\]\[-{\frac
{\phi^{2}\left (\left ({\frac {\partial ^{2}}{\partial {x}^{2}}}
\phi\right ){\frac {\partial ^{2}}{\partial {y}^{2}}}\phi- \left
({\frac {\partial ^{2}}{\partial x\partial y}}\phi\right )^
{2}\right )}{\left (1-\left ({\frac {\partial }{\partial x}}\phi
\right )^{2}-\left ({\frac {\partial }{\partial y}}\phi\right )^{
2}\right )^{2}}}-\left (1-\left ({\frac {\partial }{\partial
x}}\phi\right )^{2}-\left ({\frac {\partial }{\partial
y}}\phi\right )^{2}\right )^{-1}+1.
\end{equation}

\medskip
\subsection{Method of solution}

 For solutions of the Monge-Ampere equation
 $$F(x,y,f_x,f_y,f_{xx},f_{xy},f_{yy})=0$$ we use the method of solution of
the p.d.e.'s  described first in \cite{heav} and developed later
in \cite{dual} .

      This method allow us to construct particular solutions of the partial nonlinear
       differential equation
\begin{equation}\label{Dr10}
F(x,y,z,f_x,f_y,f_z,f_{xx},f_{xy},f_{xz},f_{yy},f_{yz},f_{xxx},f_{xyy},f_{xxy},..)=0
 \end{equation}
    with the help of transformation of the function and corresponding variables.

     Essence of the method consists in a following presentation of the functions and
     variables
\begin{equation}\label{Dr11}
f(x,y,z)\rightarrow u(x,t,z),\quad y \rightarrow v(x,t,z),\quad
f_x\rightarrow u_x-\frac{u_t}{v_t}v_x,\]\[ f_z\rightarrow
u_z-\frac{u_t}{v_t}v_z,\quad f_y \rightarrow \frac{u_t}{v_t},
\quad f_{yy} \rightarrow \frac{(\frac{u_t}{v_t})_t}{v_t}, \quad
f_{xy} \rightarrow \frac{(u_x-\frac{u_t}{v_t}v_x)_t}{v_t},...
\end{equation}
where variable $t$ is considered as parameter.

  Remark that conditions of the type
   \[
   f_{xy}=f_{yx},\quad f_{xz}=f_{zx}...
   \]
are fulfilled at the such type of presentation.

  In result instead of equation (\ref{Dr10}) one get the
  relation between the new variables $u(x,t,z)$,~ $v(x,t,z)$ and
  their partial derivatives
\begin{equation}\label{Dr12}
\Psi(u,v,u_x,u_z,u_t,v_x,v_z,v_t...)=0.
  \end{equation}

    This relation coincides with initial  p.d.e at the condition $v(x,t,z)=t$
    and lead to the new p.d.e
  \begin{equation}\label{Dr13}
     \Phi(\omega,\omega_x,\omega_t,\omega_{xx},\omega_{xt},\omega_{tt},...)=0
    \end{equation}
     when the functions $u(x,t,s)=F(\omega(x,t,z),\omega_t...)$
    and $v(x,t,s)=\Phi(\omega(x,t,z),\omega_t...)$ are expressed through the auxiliary function $\omega(x,t,s)$.

     Remark that there are a various means to reduce the relation (\ref{Dr12}) into
    the partial differential equation.

    In a some cases the solution of  equation (\ref{Dr13}) is a more simple
    problem than solution of equation (\ref{Dr10}).
\begin{rem}

  As the example we consider the  Monge-Ampere equation
    \[
   \left ({\frac {\partial ^{2}}{\partial {x}^{2}}}f(x,y)\right ){\frac {
\partial ^{2}}{\partial {y}^{2}}}f(x,y)-\left ({\frac {\partial ^{2}}{
\partial x\partial y}}f(x,y)\right )^{2}+1
=0.
\]

    After the $(u,v)$-transformation
    \[
    v(x,t)=\left ({\frac {\partial
}{\partial t}}\omega(x,t)\right )t- \omega(x,t),
\]
\[
u(x,t)=\left ({\frac {\partial }{\partial t}}\omega(x,t)\right )
\]
it takes the form of linear equation
\[
{t}^{4}{\frac {\partial ^{2}}{\partial {t}^{2}}}\omega(x,t)-{\frac
{
\partial ^{2}}{\partial {x}^{2}}}\omega(x,t)
=0.
\]
with general solution
\[
\omega(x,t)=t\left ({\it \_F1}(-{\frac {tx-1}{t}})+{\it
\_F2}({\frac { tx+1}{t}})\right ) ,
\]
depending from two arbitrary functions.

     Choice of the functions ${\it \_Fi}$ and elimination of the parameter $t$ from
     corresponding relations lead to the function $f(x,y)$
     satisfying the Monge-Ampere equation.
\end{rem}

\subsection{Congruence of cycles of zero curvature}

   In the case $K=0$ we get the equation
 \begin{equation}\label{Dr14}
\phi(x,y){\frac {\partial ^{2}}{\partial
{x}^{2}}}\phi(x,y)-\phi(x,y) \left ({\frac {\partial
^{2}}{\partial {x}^{2}}}\phi(x,y)\right ) \left ({\frac {\partial
}{\partial y}}\phi(x,y)\right )^{2}+\]\[2\,\phi(x, y)\left ({\frac
{\partial }{\partial y}}\phi(x,y)\right )\left ({ \frac {\partial
}{\partial x}}\phi(x,y)\right ){\frac {\partial ^{2}}{
\partial x\partial y}}\phi(x,y)+\phi(x,y){\frac {\partial ^{2}}{
\partial {y}^{2}}}\phi(x,y)-\]\[-\phi(x,y)\left ({\frac {\partial ^{2}}{
\partial {y}^{2}}}\phi(x,y)\right )\left ({\frac {\partial }{\partial
x}}\phi(x,y)\right )^{2}-\left (\phi(x,y)\right )^{2}\left ({\frac
{
\partial ^{2}}{\partial {x}^{2}}}\phi(x,y)\right ){\frac {\partial ^{2
}}{\partial {y}^{2}}}\phi(x,y)+\]\[+\left (\phi(x,y)\right
)^{2}\left ({ \frac {\partial ^{2}}{\partial x\partial
y}}\phi(x,y)\right )^{2}- \left ({\frac {\partial }{\partial
x}}\phi(x,y)\right )^{2}-\left ({ \frac {\partial }{\partial
y}}\phi(x,y)\right )^{2}+\]\[+\left ({\frac {
\partial }{\partial x}}\phi(x,y)\right )^{4}+2\,\left ({\frac {
\partial }{\partial x}}\phi(x,y)\right )^{2}\left ({\frac {\partial }{
\partial y}}\phi(x,y)\right )^{2}+\left ({\frac {\partial }{\partial y
}}\phi(x,y)\right )^{4}=0
\end{equation}

    After applying of the (u,v)-transformation at this  equation
 is reduced to the relation
 \[\left ({\frac {\partial }{\partial t}}u(x,t)\right )^{4}-4\,\left ({
\frac {\partial }{\partial x}}u(x,t)\right )\left ({\frac
{\partial }{
\partial t}}v(x,t)\right )\left ({\frac {\partial }{\partial t}}u(x,t)
\right )^{3}\left ({\frac {\partial }{\partial x}}v(x,t)\right
)^{3}-\]\[-4 \,\left ({\frac {\partial }{\partial x}}u(x,t)\right
)^{3}\left ({ \frac {\partial }{\partial t}}v(x,t)\right
)^{3}\left ({\frac {
\partial }{\partial t}}u(x,t)\right ){\frac {\partial }{\partial x}}v(
x,t)+\]\[+6\,\left ({\frac {\partial }{\partial x}}u(x,t)\right
)^{2}\left ({\frac {\partial }{\partial t}}v(x,t)\right )^{2}\left
({\frac {
\partial }{\partial t}}u(x,t)\right )^{2}\left ({\frac {\partial }{
\partial x}}v(x,t)\right )^{2}-\]\[-u(x,t)\left ({\frac {\partial }{
\partial t}}u(x,t)\right )^{2}\left ({\frac {\partial }{\partial t}}v(
x,t)\right )^{2}{\frac {\partial ^{2}}{\partial
{x}^{2}}}u(x,t)+u(x,t) \left ({\frac {\partial }{\partial
t}}v(x,t)\right )^{2}\left ({\frac {\partial ^{2}}{\partial
{t}^{2}}}u(x,t)\right )\left ({\frac {
\partial }{\partial x}}v(x,t)\right )^{2}-\]\[-\left (u(x,t)\right )^{2}
\left ({\frac {\partial }{\partial t}}u(x,t)\right )^{2}\left
({\frac {\partial ^{2}}{\partial {x}^{2}}}v(x,t)\right ){\frac
{\partial ^{2}} {\partial {t}^{2}}}v(x,t)+2\,\left ({\frac
{\partial }{\partial t}}u(x ,t)\right )^{4}\left ({\frac {\partial
}{\partial x}}v(x,t)\right )^{2 }-\]\[-\left ({\frac {\partial
}{\partial t}}u(x,t)\right )^{2}\left ({ \frac {\partial
}{\partial t}}v(x,t)\right )^{2}-2\,\left (u(x,t) \right
)^{2}\left ({\frac {\partial ^{2}}{\partial t\partial x}}u(x,t)
\right )\left ({\frac {\partial }{\partial t}}v(x,t)\right )\left
({ \frac {\partial }{\partial t}}u(x,t)\right ){\frac {\partial
^{2}}{
\partial t\partial x}}v(x,t)+\]\[+\left (u(x,t)\right )^{2}\left ({\frac {
\partial }{\partial t}}u(x,t)\right )\left ({\frac {\partial ^{2}}{
\partial {x}^{2}}}v(x,t)\right )\left ({\frac {\partial }{\partial t}}
v(x,t)\right ){\frac {\partial ^{2}}{\partial
{t}^{2}}}u(x,t)+\]\[+\left (u (x,t)\right )^{2}\left ({\frac
{\partial ^{2}}{\partial {x}^{2}}}u(x,t )\right )\left ({\frac
{\partial }{\partial t}}v(x,t)\right )\left ({ \frac {\partial
}{\partial t}}u(x,t)\right ){\frac {\partial ^{2}}{
\partial {t}^{2}}}v(x,t)+\left (u(x,t)\right )^{2}\left ({\frac {
\partial }{\partial t}}u(x,t)\right )^{2}\left ({\frac {\partial ^{2}}
{\partial t\partial x}}v(x,t)\right )^{2}-\]\[-u(x,t)\left ({\frac
{
\partial }{\partial t}}v(x,t)\right )^{2}\left ({\frac {\partial ^{2}}
{\partial {t}^{2}}}u(x,t)\right )\left ({\frac {\partial
}{\partial x} }u(x,t)\right )^{2}-u(x,t)\left ({\frac {\partial
}{\partial t}}v(x,t) \right )^{3}\left ({\frac {\partial
}{\partial t}}u(x,t)\right ){ \frac {\partial ^{2}}{\partial
{x}^{2}}}v(x,t)-\]\[-2\,u(x,t)\left ({\frac {\partial }{\partial
t}}v(x,t)\right )^{3}\left ({\frac {\partial ^{2} }{\partial
t\partial x}}u(x,t)\right ){\frac {\partial }{\partial x}}v
(x,t)-4\,\left ({\frac {\partial }{\partial t}}u(x,t)\right )^{3}
\left ({\frac {\partial }{\partial t}}v(x,t)\right )\left ({\frac
{
\partial }{\partial x}}u(x,t)\right ){\frac {\partial }{\partial x}}v(
x,t)-\]\[-u(x,t)\left ({\frac {\partial }{\partial t}}v(x,t)\right
)\left ( {\frac {\partial }{\partial t}}u(x,t)\right ){\frac
{\partial ^{2}}{
\partial {t}^{2}}}v(x,t)-u(x,t)\left ({\frac {\partial }{\partial t}}v
(x,t)\right )\left ({\frac {\partial }{\partial t}}u(x,t)\right )
\left ({\frac {\partial }{\partial x}}v(x,t)\right )^{2}{\frac {
\partial ^{2}}{\partial {t}^{2}}}v(x,t)+\]\[+2\,u(x,t)\left ({\frac {
\partial }{\partial t}}u(x,t)\right )\left ({\frac {\partial }{
\partial t}}v(x,t)\right )^{2}\left ({\frac {\partial }{\partial x}}u(
x,t)\right ){\frac {\partial ^{2}}{\partial t\partial
x}}u(x,t)+\]\[+2\,u(x ,t)\left ({\frac {\partial }{\partial
t}}v(x,t)\right )^{2}\left ({ \frac {\partial }{\partial
t}}u(x,t)\right )\left ({\frac {\partial }{
\partial x}}v(x,t)\right ){\frac {\partial ^{2}}{\partial t\partial x}
}v(x,t)-\]\[-2\,u(x,t)\left ({\frac {\partial }{\partial
t}}u(x,t)\right )^ {2}\left ({\frac {\partial }{\partial
t}}v(x,t)\right )\left ({\frac {
\partial }{\partial x}}u(x,t)\right ){\frac {\partial ^{2}}{\partial t
\partial x}}v(x,t)+\]\[+u(x,t)\left ({\frac {\partial }{\partial t}}v(x,t)
\right )\left ({\frac {\partial }{\partial t}}u(x,t)\right )\left
({ \frac {\partial ^{2}}{\partial {t}^{2}}}v(x,t)\right )\left
({\frac {
\partial }{\partial x}}u(x,t)\right )^{2}-\]\[-\left (u(x,t)\right )^{2}
\left ({\frac {\partial ^{2}}{\partial {x}^{2}}}u(x,t)\right
)\left ({ \frac {\partial }{\partial t}}v(x,t)\right )^{2}{\frac
{\partial ^{2}} {\partial {t}^{2}}}u(x,t)+2\,\left ({\frac
{\partial }{\partial t}}v(x ,t)\right )^{3}\left ({\frac {\partial
}{\partial x}}u(x,t)\right ) \left ({\frac {\partial }{\partial
t}}u(x,t)\right ){\frac {\partial } {\partial
x}}v(x,t)+\]\[+u(x,t)\left ({\frac {\partial }{\partial t}}u(x,t)
\right )^{3}\left ({\frac {\partial }{\partial t}}v(x,t)\right ){
\frac {\partial ^{2}}{\partial {x}^{2}}}v(x,t)+\left ({\frac {
\partial }{\partial t}}u(x,t)\right )^{4}\left ({\frac {\partial }{
\partial x}}v(x,t)\right )^{4}+\]\[+\left ({\frac {\partial }{\partial x}}u
(x,t)\right )^{4}\left ({\frac {\partial }{\partial
t}}v(x,t)\right )^ {4}-\left ({\frac {\partial }{\partial
t}}v(x,t)\right )^{4}\left ({ \frac {\partial }{\partial
x}}u(x,t)\right )^{2}-\]\[-\left ({\frac {
\partial }{\partial t}}v(x,t)\right )^{2}\left ({\frac {\partial }{
\partial t}}u(x,t)\right )^{2}\left ({\frac {\partial }{\partial x}}v(
x,t)\right )^{2}+2\,\left ({\frac {\partial }{\partial t}}u(x,t)
\right )^{2}\left ({\frac {\partial }{\partial t}}v(x,t)\right
)^{2} \left ({\frac {\partial }{\partial x}}u(x,t)\right
)^{2}+\]\[+u(x,t)\left ( {\frac {\partial }{\partial
t}}v(x,t)\right )^{2}{\frac {\partial ^{2} }{\partial
{t}^{2}}}u(x,t)+u(x,t)\left ({\frac {\partial }{\partial t}
}v(x,t)\right )^{4}{\frac {\partial ^{2}}{\partial
{x}^{2}}}u(x,t)+ \]\[+\left (u(x,t)\right )^{2}\left ({\frac
{\partial ^{2}}{\partial t
\partial x}}u(x,t)\right )^{2}\left ({\frac {\partial }{\partial t}}v(
x,t)\right )^{2} =0.
\]

From here after the choice of the functions $u$ and $v$ in the
form
\[
v(x,t)=t{\frac {\partial }{\partial t}}\omega(x,t)-\omega(x,t)
,\quad u(x,t)={\frac {\partial }{\partial t}}\omega(x,t)
\]
 we find the equation
\[
-\left ({\frac {\partial ^{2}}{\partial {t}^{2}}}\omega(x,t)\right
){t }^{2}\left ({\frac {\partial }{\partial x}}\omega(x,t)\right
)^{2}-t{ \frac {\partial }{\partial t}}\omega(x,t)-\left ({\frac
{\partial ^{2} }{\partial {t}^{2}}}\omega(x,t)\right )\left
({\frac {\partial }{
\partial t}}\omega(x,t)\right )t{\frac {\partial ^{2}}{\partial {x}^{2
}}}\omega(x,t)+\]\[+\left ({\frac {\partial ^{2}}{\partial
{t}^{2}}}\omega( x,t)\right )\left ({\frac {\partial }{\partial
t}}\omega(x,t)\right ){ t}^{3}{\frac {\partial ^{2}}{\partial
{x}^{2}}}\omega(x,t)+\left ({ \frac {\partial }{\partial
t}}\omega(x,t)\right )t\left ({\frac {
\partial ^{2}}{\partial t\partial x}}\omega(x,t)\right )^{2}-\]\[-\left ({
\frac {\partial }{\partial t}}\omega(x,t)\right ){t}^{3}\left
({\frac {\partial ^{2}}{\partial t\partial x}}\omega(x,t)\right
)^{2}-\left ({ \frac {\partial }{\partial t}}\omega(x,t)\right
)t\left ({\frac {
\partial }{\partial x}}\omega(x,t)\right )^{2}+\]\[+{\frac {\partial ^{2}}{
\partial {t}^{2}}}\omega(x,t)-\left ({\frac {\partial ^{2}}{\partial {
t}^{2}}}\omega(x,t)\right ){t}^{2}+\left ({\frac {\partial ^{2}}{
\partial {t}^{2}}}\omega(x,t)\right )\left ({\frac {\partial }{
\partial x}}\omega(x,t)\right )^{4}+\]\[+2\,\left ({\frac {\partial }{
\partial x}}\omega(x,t)\right )^{2}{\frac {\partial ^{2}}{\partial {t}
^{2}}}\omega(x,t)+2\,\left ({\frac {\partial }{\partial
t}}\omega(x,t) \right ){t}^{2}\left ({\frac {\partial
^{2}}{\partial t\partial x}} \omega(x,t)\right ){\frac {\partial
}{\partial x}}\omega(x,t)+\]\[+\left ({ \frac {\partial }{\partial
t}}\omega(x,t)\right )^{2}{\frac {\partial ^{2}}{\partial
{x}^{2}}}\omega(x,t) =0
\]
having the particular solution
\[\omega(x,t)=A(t)+x,
\]
where
\[
-2\,\left ({\frac {d^{2}}{d{t}^{2}}}A(t)\right ){t}^{2}-2\,t{\frac
{d} {dt}}A(t)+4\,{\frac {d^{2}}{d{t}^{2}}}A(t)=0.
\]

   General solution of this equation is
   \[
   A(t)={\it \_C1}+{\it \_C2}\,\ln (t+\sqrt {{t}^{2}-2})
\]

   Now elimination of the variable $t$ from the relations
\[
y\sqrt {{t}^{2}-2}-t{\it \_C2}+{\it \_C1}\,\sqrt {{t}^{2}-2}+{\it \_C2
}\,\ln (t+\sqrt {{t}^{2}-2})\sqrt {{t}^{2}-2}+x\sqrt {{t}^{2}-2}
=0,
\]
and \[ \phi(x,y)\sqrt {{t}^{2}-2}-{\it \_C2}=0
\]
give us the function $\phi(x,y)$ defined from the equation
\[
y-\sqrt {2\,\left (\phi(x,y)\right )^{2}+1}+{\it \_C1}+\ln ({\frac
{ \sqrt {2\,\left (\phi(x,y)\right )^{2}+1}+1}{\phi(x,y)}})+x=0
\]
satisfying the equation (\ref{Dr14}).

\subsection{Congruence of positive constant curvature}

In the case $K=1$ from (\ref{dr:eq3}) we get the equation
 \begin{equation}\label{Dr15}
\phi(x,y){\frac {\partial ^{2}}{\partial
{x}^{2}}}\phi(x,y)-\phi(x,y) \left ({\frac {\partial
^{2}}{\partial {x}^{2}}}\phi(x,y)\right ) \left ({\frac {\partial
}{\partial y}}\phi(x,y)\right )^{2}+\]\[+2\,\phi(x, y)\left
({\frac {\partial }{\partial y}}\phi(x,y)\right )\left ({ \frac
{\partial }{\partial x}}\phi(x,y)\right ){\frac {\partial ^{2}}{
\partial x\partial y}}\phi(x,y)+\]\[+\phi(x,y){\frac {\partial ^{2}}{
\partial {y}^{2}}}\phi(x,y)-\phi(x,y)\left ({\frac {\partial ^{2}}{
\partial {y}^{2}}}\phi(x,y)\right )\left ({\frac {\partial }{\partial
x}}\phi(x,y)\right )^{2}-\]\[-\left (\phi(x,y)\right )^{2}\left
({\frac {
\partial ^{2}}{\partial {x}^{2}}}\phi(x,y)\right ){\frac {\partial ^{2
}}{\partial {y}^{2}}}\phi(x,y)+\left (\phi(x,y)\right )^{2}\left
({ \frac {\partial ^{2}}{\partial x\partial y}}\phi(x,y)\right
)^{2}-1+\]\[+ \left ({\frac {\partial }{\partial
x}}\phi(x,y)\right )^{2}+\left ({ \frac {\partial }{\partial
y}}\phi(x,y)\right )^{2}=0.
\end{equation}

     The ($u,v$)-transformation with condition
\[
v(x,t)=t{\frac {\partial }{\partial t}}\omega(x,t)-\omega(x,t)
,\quad u(x,t)={\frac {\partial }{\partial t}}\omega(x,t)
\]
lead to the equation
\begin{equation}\label{Dr16}
\left ({\frac {\partial }{\partial t}}\omega(x,t)\right )^{2}{\frac {
\partial ^{2}}{\partial {x}^{2}}}\omega(x,t)+\left ({\frac {\partial ^
{2}}{\partial {t}^{2}}}\omega(x,t)\right ){t}^{2}-\left ({\frac {
\partial ^{2}}{\partial {t}^{2}}}\omega(x,t)\right ){t}^{4}+\]\[+2\,\left (
{\frac {\partial }{\partial t}}\omega(x,t)\right ){t}^{2}\left ({
\frac {\partial ^{2}}{\partial t\partial x}}\omega(x,t)\right
){\frac {\partial }{\partial x}}\omega(x,t)+\left ({\frac
{\partial ^{2}}{
\partial {t}^{2}}}\omega(x,t)\right )\left ({\frac {\partial }{
\partial t}}\omega(x,t)\right ){t}^{3}{\frac {\partial ^{2}}{\partial
{x}^{2}}}\omega(x,t)-\]\[-\left ({\frac {\partial ^{2}}{\partial
{t}^{2}}} \omega(x,t)\right )\left ({\frac {\partial }{\partial
t}}\omega(x,t) \right )t{\frac {\partial ^{2}}{\partial
{x}^{2}}}\omega(x,t)-\left ({ \frac {\partial }{\partial
t}}\omega(x,t)\right ){t}^{3}\left ({\frac {\partial
^{2}}{\partial t\partial x}}\omega(x,t)\right )^{2}+\]\[+\left ({
\frac {\partial }{\partial t}}\omega(x,t)\right )t\left ({\frac {
\partial ^{2}}{\partial t\partial x}}\omega(x,t)\right )^{2}+\left ({
\frac {\partial ^{2}}{\partial {t}^{2}}}\omega(x,t)\right ){t}^{2}
\left ({\frac {\partial }{\partial x}}\omega(x,t)\right
)^{2}-t{\frac {\partial }{\partial t}}\omega(x,t)-\]\[-\left
({\frac {\partial }{
\partial t}}\omega(x,t)\right )t\left ({\frac {\partial }{\partial x}}
\omega(x,t)\right )^{2}=0. \end{equation}

     This equation admits particular solution in the form
    \[
    \omega(x,t)=A(t)+x
\]
where the function $A(t)$ is solution of equation
\[2\,\left
({\frac {d^{2}}{d{t}^{2}}}A(t)\right ){t}^{2}-\left ({\frac {
d^{2}}{d{t}^{2}}}A(t)\right ){t}^{4}-2\,t{\frac {d}{dt}}A(t)=0.
\]

    So we get
 \[
 A(t)={\it \_C1}+{\it \_C2}\,\sqrt {-2+{t}^{2}}
\]

     Now after elimination of the parameter $t$ from the relations
     \[y\sqrt {-2+{t}^{2}}+{\it \_C1}\,\sqrt {-2+{t}^{2}}-2\,{\it \_C2}+x
\sqrt {-2+{t}^{2}}=0
\]
and
\[\phi(x,y)\sqrt {{t}^{2}-2}-{\it \_C2}
=0
\]
we obtain the simplest solution of the equation (\ref{Dr15}).
\[
\phi(x,y)=1/2\,\sqrt {2\,{y}^{2}+4\,y{\it \_C1}+4\,yx+2\,{{\it
\_C1}}^{2}+4\,{ \it \_C1}\,x+2\,{x}^{2}+4\,{{\it \_C2}}^{2}}.
\]

    Moore complicated solutions of the equation (\ref{Dr16}) in the
    form
\[ \omega(x,t)=A(t)+B(t)x
\]
lead to the conditions
\[
B(t)=-\sqrt {{t}^{2}-1},
\]
and
 \[
A(t)
\]
is arbitrary function.

 In result elimination of the parameter
$t$ from the relations
\[
y\sqrt {{t}^{2}-1}-t\left ({\frac {d}{dt}}A(t)\right )\sqrt
{{t}^{2}-1 }+A(t)\sqrt {{t}^{2}-1}+x=0,
\]
and
\[\phi(x,y)\sqrt {{t}^{2}-1}-\left ({\frac {d}{dt}}A(t)\right )\sqrt {{t
}^{2}-1}+tx=0
\]
with a given function $A(t)$ we get the solution of the equation
(\ref{Dr15}) dependent from choice of arbitrary function.

As example in the  case
\[
A(t)=\frac{1}{t}
\]
we find the solution of the equation (\ref{Dr15}) in the form
\[16\,\left (\phi(x,y)\right )^{4}+\left (8\,{y}^{2}-8\,{x}^{2}-32
\right )\left (\phi(x,y)\right )^{3}+\left
(-32\,{y}^{2}+16-8\,{x}^{2}
+{x}^{4}+{y}^{4}+2\,{y}^{2}{x}^{2}\right )\left (\phi(x,y)\right
)^{2} +\]\[+\left
(8\,{y}^{2}-10\,{y}^{4}+8\,{x}^{4}-2\,{y}^{2}{x}^{2}+32\,{x}^{2
}\right
)\phi(x,y)-{y}^{6}-{x}^{6}+20\,{y}^{2}{x}^{2}-8\,{x}^{4}-3\,{y
}^{2}{x}^{4}-16\,{x}^{2}-3\,{y}^{4}{x}^{2}+{y}^{4}=0.
\]
\subsection{Congruence of negative constant curvature}

In the case $K=-1$ from (\ref{dr:eq3}) we get the equation
 \begin{equation}\label{Dr17}
 \phi(x,y){\frac {\partial ^{2}}{\partial
{x}^{2}}}\phi(x,y)-\phi(x,y) \left ({\frac {\partial
^{2}}{\partial {x}^{2}}}\phi(x,y)\right ) \left ({\frac {\partial
}{\partial y}}\phi(x,y)\right )^{2}+\]\[+2\,\phi(x, y)\left
({\frac {\partial }{\partial y}}\phi(x,y)\right )\left ({ \frac
{\partial }{\partial x}}\phi(x,y)\right ){\frac {\partial ^{2}}{
\partial x\partial y}}\phi(x,y)+\phi(x,y){\frac {\partial ^{2}}{
\partial {y}^{2}}}\phi(x,y)-\]\[-\phi(x,y)\left ({\frac {\partial ^{2}}{
\partial {y}^{2}}}\phi(x,y)\right )\left ({\frac {\partial }{\partial
x}}\phi(x,y)\right )^{2}-\left (\phi(x,y)\right )^{2}\left ({\frac
{
\partial ^{2}}{\partial {x}^{2}}}\phi(x,y)\right ){\frac {\partial ^{2
}}{\partial {y}^{2}}}\phi(x,y)+\]\[+\left (\phi(x,y)\right
)^{2}\left ({ \frac {\partial ^{2}}{\partial x\partial
y}}\phi(x,y)\right )^{2}+1-3 \,\left ({\frac {\partial }{\partial
x}}\phi(x,y)\right )^{2}-3\, \left ({\frac {\partial }{\partial
y}}\phi(x,y)\right )^{2}+\]\[+2\,\left ( {\frac {\partial
}{\partial x}}\phi(x,y)\right )^{4}+4\,\left ({\frac {\partial
}{\partial x}}\phi(x,y)\right )^{2}\left ({\frac {\partial }
{\partial y}}\phi(x,y)\right )^{2}+2\,\left ({\frac {\partial }{
\partial y}}\phi(x,y)\right )^{4}=0.
\end{equation}

     The ($u,v$)-transformation with condition
\[
u(x,t)=t{\frac {\partial }{\partial t}}\omega(x,t)-\omega(x,t)
,\quad v(x,t)={\frac {\partial }{\partial t}}\omega(x,t)
\]
lead to the equation
\begin{equation}\label{Dr18}
\left ({\frac {\partial }{\partial t}}\omega(x,t)\right
)^{2}{\frac {
\partial ^{2}}{\partial {x}^{2}}}\omega(x,t)+\left ({\frac {\partial ^
{2}}{\partial {t}^{2}}}\omega(x,t)\right ){t}^{2}-\left ({\frac {
\partial ^{2}}{\partial {t}^{2}}}\omega(x,t)\right ){t}^{4}+\]\[+2\,\left (
{\frac {\partial }{\partial t}}\omega(x,t)\right ){t}^{2}\left ({
\frac {\partial ^{2}}{\partial t\partial x}}\omega(x,t)\right
){\frac {\partial }{\partial x}}\omega(x,t)+\left ({\frac
{\partial ^{2}}{
\partial {t}^{2}}}\omega(x,t)\right )\left ({\frac {\partial }{
\partial t}}\omega(x,t)\right ){t}^{3}{\frac {\partial ^{2}}{\partial
{x}^{2}}}\omega(x,t)-\]\[-\left ({\frac {\partial ^{2}}{\partial
{t}^{2}}} \omega(x,t)\right )\left ({\frac {\partial }{\partial
t}}\omega(x,t) \right )t{\frac {\partial ^{2}}{\partial
{x}^{2}}}\omega(x,t)-\left ({ \frac {\partial }{\partial
t}}\omega(x,t)\right ){t}^{3}\left ({\frac {\partial
^{2}}{\partial t\partial x}}\omega(x,t)\right )^{2}+\]\[+\left ({
\frac {\partial }{\partial t}}\omega(x,t)\right )t\left ({\frac {
\partial ^{2}}{\partial t\partial x}}\omega(x,t)\right )^{2}+\left ({
\frac {\partial ^{2}}{\partial {t}^{2}}}\omega(x,t)\right ){t}^{2}
\left ({\frac {\partial }{\partial x}}\omega(x,t)\right
)^{2}-t{\frac {\partial }{\partial t}}\omega(x,t)-\]\[-\left
({\frac {\partial }{
\partial t}}\omega(x,t)\right )t\left ({\frac {\partial }{\partial x}}
\omega(x,t)\right )^{2}=0.
 \end{equation}

     This equation admits the particular solution
         \[
\omega(x,t)=\sqrt {-{t}^{2}+1}x+A(t)
\]
with arbitrary function $A(t)$.

    In particular case after elimination of the parameter $t$
    from the relations
    \[
    y\sqrt {-{t}^{2}+1}+xt-2\,t\sqrt {-{t}^{2}+1}=0
\]
and
\[
\phi(x,y)\sqrt {-{t}^{2}+1}-{t}^{2}\sqrt {-{t}^{2}+1}+x=0
\]
we find the solution of the equation (\ref{Dr17}) in the form
\[
-16\,\left (\phi(x,y)\right )^{4}+\left (-32+8\,{y}^{2}-8\,{x}^{2}
\right )\left (\phi(x,y)\right )^{3}+\]\[+\left
(-2\,{y}^{2}{x}^{2}+32\,{y}
^{2}-16-{y}^{4}+8\,{x}^{2}-{x}^{4}\right )\left (\phi(x,y)\right
)^{2} +\]\[+\left
(-2\,{y}^{2}{x}^{2}-10\,{y}^{4}+32\,{x}^{2}+8\,{x}^{4}+8\,{y}^{
2}\right
)\phi(x,y)-\]\[-{y}^{4}+16\,{x}^{2}+3\,{x}^{4}{y}^{2}+8\,{x}^{4}+3
\,{y}^{4}{x}^{2}+{x}^{6}+{y}^{6}-20\,{y}^{2}{x}^{2}=0
\]
\section{Geodesic equations}

    The geodesic of the metric (\ref{dr:eq1}) are equivalent to the equation
    \[
    {\frac {d^{2}}{d{x}^{2}}}y(x)+{\it a\_1}(x,y)\left ({\frac {d}{dx}}y(x
)\right )^{3}+3\,{\it a\_2}(x,y)\left ({\frac {d}{dx}}y(x)\right
)^{2} +3\,{\it a\_3}(x,y){\frac {d}{dx}}y(x)+{\it a\_4}(x,y)=0,
 \]
where
\[
\it a\_1(x,y)={\frac {\left ({\frac {\partial }{\partial
x}}\phi(x,y)\right )\left ( \left ({\frac {\partial ^{2}}{\partial
{y}^{2}}}\phi(x,y)\right )\phi( x,y)-1+\left ({\frac {\partial
}{\partial y}}\phi(x,y)\right )^{2} \right )}{\phi(x,y)\left
(-1+\left ({\frac {\partial }{\partial y}} \phi(x,y)\right
)^{2}+\left ({\frac {\partial }{\partial x}}\phi(x,y) \right
)^{2}\right )}},
\]
\[
3\,{\it a\_2}(x,y)={\frac {\left ({\frac {\partial }{\partial
y}}\phi( x,y)\right ){\frac {\partial ^{2}}{\partial
{y}^{2}}}\phi(x,y)}{-1+ \left ({\frac {\partial }{\partial
y}}\phi(x,y)\right )^{2}+\left ({ \frac {\partial }{\partial
x}}\phi(x,y)\right )^{2}}}+2\,{\frac { \left ({\frac {\partial
}{\partial x}}\phi(x,y)\right ){\frac {
\partial ^{2}}{\partial x\partial y}}\phi(x,y)}{-1+\left ({\frac {
\partial }{\partial y}}\phi(x,y)\right )^{2}+\left ({\frac {\partial }
{\partial x}}\phi(x,y)\right )^{2}}}+\]\[+{\frac {-3\,\left
({\frac {
\partial }{\partial y}}\phi(x,y)\right )^{3}-2\,\left ({\frac {
\partial }{\partial x}}\phi(x,y)\right )^{2}{\frac {\partial }{
\partial y}}\phi(x,y)+3\,{\frac {\partial }{\partial y}}\phi(x,y)}{
\phi(x,y)\left (-1+\left ({\frac {\partial }{\partial y}}\phi(x,y)
\right )^{2}+\left ({\frac {\partial }{\partial x}}\phi(x,y)\right
)^{ 2}\right )}},
\]
\[
3\,{\it a\_3}(x,y)={\frac {\left ({\frac {\partial }{\partial
x}}\phi( x,y)\right ){\frac {\partial ^{2}}{\partial
{x}^{2}}}\phi(x,y)}{-1+ \left ({\frac {\partial }{\partial
y}}\phi(x,y)\right )^{2}+\left ({ \frac {\partial }{\partial
x}}\phi(x,y)\right )^{2}}}+2\,{\frac { \left ({\frac {\partial
}{\partial y}}\phi(x,y)\right ){\frac {
\partial ^{2}}{\partial x\partial y}}\phi(x,y)}{-1+\left ({\frac {
\partial }{\partial y}}\phi(x,y)\right )^{2}+\left ({\frac {\partial }
{\partial x}}\phi(x,y)\right )^{2}}}+\]\[+{\frac {-2\,\left
({\frac {
\partial }{\partial x}}\phi(x,y)\right )\left ({\frac {\partial }{
\partial y}}\phi(x,y)\right )^{2}+3\,{\frac {\partial }{\partial x}}
\phi(x,y)-3\,\left ({\frac {\partial }{\partial x}}\phi(x,y)\right
)^{ 3}}{\phi(x,y)\left (-1+\left ({\frac {\partial }{\partial
y}}\phi(x,y) \right )^{2}+\left ({\frac {\partial }{\partial
x}}\phi(x,y)\right )^{ 2}\right )}},
\]
\[
{\it a\_4}(x,y)={\frac {\left ({\frac {\partial }{\partial
y}}\phi(x,y )\right )\left (\left ({\frac {\partial }{\partial
x}}\phi(x,y)\right )^{2}+\left ({\frac {\partial ^{2}}{\partial
{x}^{2}}}\phi(x,y)\right )\phi(x,y)-1\right )}{\phi(x,y)\left
(-1+\left ({\frac {\partial }{
\partial y}}\phi(x,y)\right )^{2}+\left ({\frac {\partial }{\partial x
}}\phi(x,y)\right )^{2}\right )}}.
\]
\section{The four-dimensional Riemann extension}

    The metric (\ref{dr:eq1}) has a following coefficients of connection
\[
\Gamma^{1}_{11}={\frac {\left ({\frac {\partial }{
\partial x}}\phi(x,y)\right )\left (1-\left ({\frac {\partial }{
\partial x}}\phi(x,y)\right )^{2}+\left ({\frac {\partial ^{2}}{
\partial {x}^{2}}}\phi(x,y)\right )\phi(x,y)-2\,\left ({\frac {
\partial }{\partial y}}\phi(x,y)\right )^{2}\right )}{\phi(x,y)\left (
-1+\left ({\frac {\partial }{\partial y}}\phi(x,y)\right
)^{2}+\left ( {\frac {\partial }{\partial x}}\phi(x,y)\right
)^{2}\right )}} ,
\]
 \[
 \Gamma^{2}_{11}={\frac {\left ({\frac {\partial }{\partial y}}\phi(x,y)\right )\left (
\left ({\frac {\partial }{\partial x}}\phi(x,y)\right )^{2}+\left
({ \frac {\partial ^{2}}{\partial {x}^{2}}}\phi(x,y)\right
)\phi(x,y)-1 \right )}{\phi(x,y)\left (-1+\left ({\frac {\partial
}{\partial y}} \phi(x,y)\right )^{2}+\left ({\frac {\partial
}{\partial x}}\phi(x,y) \right )^{2}\right )}},
\]
\[
 \Gamma^{1}_{12}={\frac {{\frac {\partial }{\partial y}}\phi(x,y)+\left ({\frac {
\partial }{\partial x}}\phi(x,y)\right )\left ({\frac {\partial ^{2}}{
\partial x\partial y}}\phi(x,y)\right )\phi(x,y)-\left ({\frac {
\partial }{\partial y}}\phi(x,y)\right )^{3}}{\phi(x,y)\left (-1+
\left ({\frac {\partial }{\partial y}}\phi(x,y)\right )^{2}+\left
({ \frac {\partial }{\partial x}}\phi(x,y)\right )^{2}\right )}},
\]
\[
 \Gamma^{2}_{12}={\frac {{\frac {\partial }{\partial x}}\phi(x,y)+\left ({\frac {
\partial ^{2}}{\partial x\partial y}}\phi(x,y)\right )\left ({\frac {
\partial }{\partial y}}\phi(x,y)\right )\phi(x,y)-\left ({\frac {
\partial }{\partial x}}\phi(x,y)\right )^{3}}{\phi(x,y)\left (-1+
\left ({\frac {\partial }{\partial y}}\phi(x,y)\right )^{2}+\left
({ \frac {\partial }{\partial x}}\phi(x,y)\right )^{2}\right )}},
\]
\[
 \Gamma^{1}_{22}={\frac {\left ({\frac {\partial }{\partial x}}\phi(x,y)\right )\left (
\left ({\frac {\partial ^{2}}{\partial {y}^{2}}}\phi(x,y)\right
)\phi( x,y)-1+\left ({\frac {\partial }{\partial
y}}\phi(x,y)\right )^{2} \right )}{\phi(x,y)\left (-1+\left
({\frac {\partial }{\partial y}} \phi(x,y)\right )^{2}+\left
({\frac {\partial }{\partial x}}\phi(x,y) \right )^{2}\right )}},
\]
\[
 \Gamma^{1}_{22}={\frac {\left ({\frac {\partial }{\partial y}}\phi(x,y)\right
)\left ( -2\,\left ({\frac {\partial }{\partial x}}\phi(x,y)\right
)^{2}+1- \left ({\frac {\partial }{\partial y}}\phi(x,y)\right
)^{2}+\left ({ \frac {\partial ^{2}}{\partial
{y}^{2}}}\phi(x,y)\right )\phi(x,y) \right )}{\phi(x,y)\left
(-1+\left ({\frac {\partial }{\partial y}} \phi(x,y)\right
)^{2}+\left ({\frac {\partial }{\partial x}}\phi(x,y) \right
)^{2}\right )}}.
\]

   Using given expressions for the connections coefficients we
   introduce a 4-dimensional Riemann space
   with the metric
   \begin{equation} \label{Dr18}
{^{4}}ds^2=\left(-2\Gamma^1_{ij}z-2\Gamma^2_{ij}t\right) dx^i
dx^j+2d x dz+2dy dt
\end{equation}
where $z$ and $t$ are an additional coordinates.

     The Riemann space constructed on such a way is called the Riemann extension
     of the base space equipped with connection \cite{weyl}.

     In explicit form the non zero components of the metric (\ref{Dr18})
     looks as
     \[
    g_{xx}=-2\,{\frac {\left ({\frac {\partial }{
\partial x}}\phi(x,y)\right )z}{\phi(x,y)\left (-1+\left ({\frac {
\partial }{\partial y}}\phi(x,y)\right )^{2}+\left ({\frac {\partial }
{\partial x}}\phi(x,y)\right )^{2}\right )}}+2\,{\frac {\left
({\frac {\partial }{\partial x}}\phi(x,y)\right
)^{3}z}{\phi(x,y)\left (-1+ \left ({\frac {\partial }{\partial
y}}\phi(x,y)\right )^{2}+\left ({ \frac {\partial }{\partial
x}}\phi(x,y)\right )^{2}\right )}}-\]\[-2\,{ \frac {\left ({\frac
{\partial }{\partial x}}\phi(x,y)\right )z{\frac {\partial
^{2}}{\partial {x}^{2}}}\phi(x,y)}{-1+\left ({\frac {
\partial }{\partial y}}\phi(x,y)\right )^{2}+\left ({\frac {\partial }
{\partial x}}\phi(x,y)\right )^{2}}}+4\,{\frac {\left ({\frac {
\partial }{\partial x}}\phi(x,y)\right )z\left ({\frac {\partial }{
\partial y}}\phi(x,y)\right )^{2}}{\phi(x,y)\left (-1+\left ({\frac {
\partial }{\partial y}}\phi(x,y)\right )^{2}+\left ({\frac {\partial }
{\partial x}}\phi(x,y)\right )^{2}\right )}}-\]\[-2\,{\frac {\left
({\frac {\partial }{\partial y}}\phi(x,y)\right )t\left ({\frac
{\partial }{
\partial x}}\phi(x,y)\right )^{2}}{\phi(x,y)\left (-1+\left ({\frac {
\partial }{\partial y}}\phi(x,y)\right )^{2}+\left ({\frac {\partial }
{\partial x}}\phi(x,y)\right )^{2}\right )}}-2\,{\frac {\left
({\frac {\partial }{\partial y}}\phi(x,y)\right )t{\frac {\partial
^{2}}{
\partial {x}^{2}}}\phi(x,y)}{-1+\left ({\frac {\partial }{\partial y}}
\phi(x,y)\right )^{2}+\left ({\frac {\partial }{\partial
x}}\phi(x,y) \right )^{2}}}+\]\[+2\,{\frac {\left ({\frac
{\partial }{\partial y}}\phi(x ,y)\right )t}{\phi(x,y)\left
(-1+\left ({\frac {\partial }{\partial y} }\phi(x,y)\right
)^{2}+\left ({\frac {\partial }{\partial x}}\phi(x,y) \right
)^{2}\right )}},
\]
\[
g_{xy}=-2\,{\frac {z{\frac {\partial }{
\partial y}}\phi(x,y)}{\phi(x,y)\left (-1+\left ({\frac {\partial }{
\partial y}}\phi(x,y)\right )^{2}+\left ({\frac {\partial }{\partial x
}}\phi(x,y)\right )^{2}\right )}}-2\,{\frac {\left ({\frac
{\partial } {\partial x}}\phi(x,y)\right )z{\frac {\partial
^{2}}{\partial x
\partial y}}\phi(x,y)}{-1+\left ({\frac {\partial }{\partial y}}\phi(x
,y)\right )^{2}+\left ({\frac {\partial }{\partial
x}}\phi(x,y)\right )^{2}}}+\]\[+2\,{\frac {z\left ({\frac
{\partial }{\partial y}}\phi(x,y) \right )^{3}}{\phi(x,y)\left
(-1+\left ({\frac {\partial }{\partial y} }\phi(x,y)\right
)^{2}+\left ({\frac {\partial }{\partial x}}\phi(x,y) \right
)^{2}\right )}}-2\,{\frac {t{\frac {\partial }{\partial x}}\phi
(x,y)}{\phi(x,y)\left (-1+\left ({\frac {\partial }{\partial
y}}\phi(x ,y)\right )^{2}+\left ({\frac {\partial }{\partial
x}}\phi(x,y)\right )^{2}\right )}}-\]\[-2\,{\frac {t\left ({\frac
{\partial ^{2}}{\partial x
\partial y}}\phi(x,y)\right ){\frac {\partial }{\partial y}}\phi(x,y)}
{-1+\left ({\frac {\partial }{\partial y}}\phi(x,y)\right
)^{2}+\left ({\frac {\partial }{\partial x}}\phi(x,y)\right
)^{2}}}+2\,{\frac {t \left ({\frac {\partial }{\partial
x}}\phi(x,y)\right )^{3}}{\phi(x,y) \left (-1+\left ({\frac
{\partial }{\partial y}}\phi(x,y)\right )^{2}+ \left ({\frac
{\partial }{\partial x}}\phi(x,y)\right )^{2}\right )}},
\]
\[g_{yy}=-2\,{\frac {\left ({\frac {\partial }{
\partial x}}\phi(x,y)\right )z{\frac {\partial ^{2}}{\partial {y}^{2}}
}\phi(x,y)}{-1+\left ({\frac {\partial }{\partial
y}}\phi(x,y)\right ) ^{2}+\left ({\frac {\partial }{\partial
x}}\phi(x,y)\right )^{2}}}+2\, {\frac {\left ({\frac {\partial
}{\partial x}}\phi(x,y)\right )z}{\phi (x,y)\left (-1+\left
({\frac {\partial }{\partial y}}\phi(x,y)\right ) ^{2}+\left
({\frac {\partial }{\partial x}}\phi(x,y)\right )^{2} \right
)}}-\]\[-2\,{\frac {\left ({\frac {\partial }{\partial
x}}\phi(x,y) \right )z\left ({\frac {\partial }{\partial
y}}\phi(x,y)\right )^{2}}{ \phi(x,y)\left (-1+\left ({\frac
{\partial }{\partial y}}\phi(x,y) \right )^{2}+\left ({\frac
{\partial }{\partial x}}\phi(x,y)\right )^{ 2}\right )}}+4\,{\frac
{\left ({\frac {\partial }{\partial y}}\phi(x,y )\right )t\left
({\frac {\partial }{\partial x}}\phi(x,y)\right )^{2}}
{\phi(x,y)\left (-1+\left ({\frac {\partial }{\partial
y}}\phi(x,y) \right )^{2}+\left ({\frac {\partial }{\partial
x}}\phi(x,y)\right )^{ 2}\right )}}-\]\[-2\,{\frac {\left ({\frac
{\partial }{\partial y}}\phi(x,y )\right )t}{\phi(x,y)\left
(-1+\left ({\frac {\partial }{\partial y}} \phi(x,y)\right
)^{2}+\left ({\frac {\partial }{\partial x}}\phi(x,y) \right
)^{2}\right )}}+2\,{\frac {\left ({\frac {\partial }{\partial y
}}\phi(x,y)\right )^{3}t}{\phi(x,y)\left (-1+\left ({\frac
{\partial } {\partial y}}\phi(x,y)\right )^{2}+\left ({\frac
{\partial }{\partial x}}\phi(x,y)\right )^{2}\right
)}}-\]\[-2\,{\frac {\left ({\frac {\partial }{\partial
y}}\phi(x,y)\right )t{\frac {\partial ^{2}}{\partial {y}^{2
}}}\phi(x,y)}{-1+\left ({\frac {\partial }{\partial y}}\phi(x,y)
\right )^{2}+\left ({\frac {\partial }{\partial x}}\phi(x,y)\right
)^{ 2}}},
\]
\[
g_{yt}=1,\quad g_{xz}=1.
\]

\begin{pr}  Riemann space with the metric (\ref{Dr18}) is a
Ricci-flat \[R_{ij}=0\] at the condition
  \begin{equation} \label{Dr19}
-2\,\left ({\frac {\partial }{\partial x}}\phi(x,y)\right
)^{2}\left ( {\frac {\partial }{\partial y}}\phi(x,y)\right
)^{2}-\left ({\frac {
\partial }{\partial x}}\phi(x,y)\right )^{4}+\left ({\frac {\partial }
{\partial y}}\phi(x,y)\right )^{2}+\left ({\frac {\partial
}{\partial x}}\phi(x,y)\right )^{2}-\]\[-2\,\left ({\frac
{\partial }{\partial x}}\phi (x,y)\right )\phi(x,y)\left ({\frac
{\partial }{\partial y}}\phi(x,y) \right ){\frac {\partial
^{2}}{\partial x\partial y}}\phi(x,y)+\left ( {\frac {\partial
^{2}}{\partial {x}^{2}}}\phi(x,y)\right )\phi(x,y) \left ({\frac
{\partial }{\partial y}}\phi(x,y)\right )^{2}+\]\[+\left ({ \frac
{\partial ^{2}}{\partial {y}^{2}}}\phi(x,y)\right )\left ({ \frac
{\partial ^{2}}{\partial {x}^{2}}}\phi(x,y)\right )\left (\phi(x
,y)\right )^{2}+\left ({\frac {\partial ^{2}}{\partial
{y}^{2}}}\phi(x ,y)\right )\left ({\frac {\partial }{\partial
x}}\phi(x,y)\right )^{2} \phi(x,y)-\]\[-\left ({\frac {\partial
^{2}}{\partial {y}^{2}}}\phi(x,y) \right )\phi(x,y)-\left ({\frac
{\partial ^{2}}{\partial {x}^{2}}}\phi (x,y)\right
)\phi(x,y)-\]\[-\left ({\frac {\partial ^{2}}{\partial x
\partial y}}\phi(x,y)\right )^{2}\left (\phi(x,y)\right )^{2}-\left ({
\frac {\partial }{\partial y}}\phi(x,y)\right )^{4}=0.
\end{equation}
\end{pr}

    Remark that  two dimensional metric (\ref{dr:eq1}) at this condition is a
    flat.

   It is important to note that the space with the metric (\ref{Dr18})
   with condition (\ref{Dr19}) does not a
   flat, the component $R_{1212}$ of its  Riemann tensor  $R_{1212}\neq0$.

   So in result of the Riemann extension of the metric (\ref{dr:eq1})
   we have got the Einstein space.

    Finally we demonstrate some additional solutions of the equation
    (\ref{Dr19}).

      The substitution
    \[
    \phi(x,y)=H(x+y)\]
into the equation (\ref{Dr19}) lead to the condition on the
function $H(x+y)=H(z)$
\[2\,\left (\mbox {D}(H)(z)\right )^{4}+\left (D^{\left (2\right )}
\right )(H)(z)H(z)-\left (\mbox {D}(H)(z)\right )^{2}=0.
\]

   From here we find
the function $H(z)$ in non explicit form
\[\sqrt {2\,\left
(H(z)\right )^{2}-{\it \_C1}}+{\it \_C1}\,\ln ({\frac {-2\,{\it
\_C1}+2\,\sqrt {-{\it \_C1}}\sqrt {2\,\left (H(z)\right )^{2
}-{\it \_C1}}}{H(z)}}){\frac {1}{\sqrt {-{\it \_C1}}}}-z-{\it
\_C2}=0.
\]

The substitution
\[\phi(x,y)=H({\frac {y}{x}})x
\]
lead to the complex solutions.

    In additive to the part $(2.2)$  we show the solutions of the
the equation (\ref{Dr19}) (or) which is connected with the
function $\omega(x,t)$ in form
\[
\omega(x,t)=A(t)+x\sqrt {{t}^{2}-1},
\]
where $A(t)$ is arbitrary function.

In particular case $A(t)=t^2$ from here is followed that the
function $\phi(x,y)$ defined from the equation
\[
-\left (\phi(x,y)\right )^{6}+\left
(3\,{x}^{2}+1+10\,y+{y}^{2}\right )\left (\phi(x,y)\right
)^{4}+\]\[+\left (-2\,y{x}^{2}-2\,{x}^{2}{y}^{2}-20
\,{x}^{2}-8\,y-8\,{y}^{3}-3\,{x}^{4}-32\,{y}^{2}\right )\left
(\phi(x, y)\right
)^{2}+\]\[+{y}^{2}{x}^{4}-8\,{x}^{4}y+16\,{y}^{4}+{x}^{6}+16\,{y}^
{2}+32\,{y}^{3}+8\,{x}^{2}{y}^{2}+32\,y{x}^{2}+16\,{x}^{2}-8\,{x}^{4}-
8\,{y}^{3}{x}^{2}=0
\]
is the solution of the equation (\ref{Dr19}).

\end{document}